\documentclass[twoside,leqno,twocolumn]{article}

\usepackage[letterpaper]{geometry}

\usepackage{ltexpprt}
\usepackage{hyperref}

\usepackage{savetrees}[subtle]
\usepackage{xcolor}
\usepackage{soul}
\usepackage{adjustbox}
\usepackage{pgfplots}
\pgfplotsset{width=10cm,compat=1.9}
\usepgfplotslibrary{external}
\usepgfplotslibrary{groupplots}
\usepackage{caption}
\usepackage{subcaption}
\usepackage{csvsimple}
\usepackage{algorithm}
\usepackage{algpseudocode}
\usepackage{svg}
\usepackage{mathtools}
\usepackage{xspace}
\usepackage{amsfonts}
\usepackage{tikz}
\usepackage{url}
\usepackage{enumerate}
\usepackage{enumitem}
\usepackage[normalem]{ulem}
\usepackage{cleveref}
\usepackage{balance}
\usepackage{makecell}
\usepackage{multirow}
\usepackage{booktabs} 

\usepackage[normalem]{ulem}

\begin{document}

%

\newcommand{\btree}{B-tree\xspace}
\newcommand{\btrees}{B-trees\xspace}
\newcommand{\lj}{learned index-based join\xspace}
\newcommand{\ljs}{learned index-based joins\xspace}
\newcommand{\chesetti}[2]{{\color{blue} #1 $\langle$\sffamily YC: #2$\rangle$}}
\newcommand{\para}[1]{\smallskip\noindent\textbf{#1.}}
\newcommand{\defn}[1]{{\textit{\textbf{\boldmath #1}}}\xspace}
\newcommand{\todo}[2]{{\color{purple} #1}} 
\newcommand{\diff}[2]{{\color{red} \sout{#1}}{\color{blue} #2}}
\newcommand{\update}[1]{{\color{blue} \sffamily }}

\title{\Large Evaluating Learned Indexes for External Memory Joins}
\author{Yuvaraj Chesetti\thanks{Northeastern University.}
\and Prashant Pandey\thanks{Northeastern University. }}

\date{}

\maketitle


\fancyfoot[R]{\scriptsize{Copyright \textcopyright\ 2025 by SIAM\\
Unauthorized reproduction of this article is prohibited}}





\begin{abstract} \small\baselineskip=9pt 

Joins are among the most time-consuming and data-intensive operations in relational query processing. Much research effort has been applied to the optimization of join processing due to their frequent execution. Recent studies have shown that CDF-based learned models can create smaller and faster indexes, accelerating in-memory joins. However, their effectiveness for external-memory joins, which are crucial for large-scale databases, remains underexplored.

This paper evaluates the impact of learned indexes on external-memory joins for both sorted and unsorted data. We compare learned index-based joins against traditional join methods such as hash joins, sort joins, and indexed nested-loop joins on real-world and simulated datasets. Additionally, we analyze learned index-based joins across multiple dimensions, including storage device types, data sorting, parallelism, constrained memory environments, and varying model error. The detailed evaluation enables us to determine the most appropriate learned index to employ for external-memory joins.

Our experiments reveal that, unlike in-memory settings, learned indexes in external-memory joins can trade off accuracy for space without significantly degrading performance. While learned indexes provide smaller index sizes and faster lookups, they perform similarly to B-trees in external-memory joins since the total amount of I/O, which dominates runtime, remains unchanged. Additionally, the construction times of learned indexes are $\sim 1000\times$ longer, and although they are $2–4\times$ smaller than the internal nodes of a B-tree, these nodes only represent 0.4\%–1\% of the data size and typically fit in main memory.

\end{abstract}
\section{Introduction}

The join operation is a fundamental operation in database management systems. Data normalization spreads data across multiple relational tables, and the join operation enables combining tuples from two or more relations based on a common attribute. Implementing join efficiently is challenging as it involves iterating through the tuples across multiple relations, incurring large amounts of disk I/Os. The join operation is often critical for the overall database performance as it is expensive and frequently executed. This is especially true for online analytics processing (OLAP) systems where the data is static and large-scale multidimensional analytical queries are frequent. This has resulted in extensive research on optimizing joins in recent decades~\cite{albutiu2012a, balkesen2013, bandle2021, blanas_design_2011,thostrup,sabek2023, krastnikov_efficient_2020}.


Recently, machine learning has significantly influenced the automation of fundamental database functions and design choices. Specifically, researchers have shown that indexes based on learned models that approximate the cumulative distribution function (CDF)~\cite{kraska_case_2018}---which is effectively the rank function for items in a dataset ---- are faster and smaller than traditional indexing data structures~\cite{marcus2020a}.
These learned structures and algorithms often outperform their traditional counterparts because they can accurately capture data trends and optimize performance for specific instances. 
A recent survey~\cite{almamun2024survey} identifies close to 100 proposed learned indexes in the last five years.

A recent benchmarking study~\cite{marcus2020a} demonstrates that learned index structures such as the PGM-Index~\cite{ferragina_pgm-index_2020}, RMI~\cite{kraska_case_2018} and RadixSpline~\cite{kipf_radixspline_2020} can deliver superior lookup performance on sorted data compared to traditional indexes. Kristo et al.~\cite{DBLP:conf/sigmod/KristoVCMK20} show that learned models can speed up sorting in main memory for integer keys. Sabek and Kraska~\cite{sabek2023} demonstrate that learned models can improve in-memory indexed nested loop joins by $5-25\%$ and can improve the partitioning step in parallel hash and sort-joins.

While recent work on in-memory benchmarks has shown promising results, the benefits of learned indexes in the external memory settings do not appear to be as clear. Lan et al.~\cite{lan_updatable_2023} adapt multiple updatable learned indexes to external memory and evaluate performance on read, scan, and write workloads. They observe that in their current iterations, none of the updatable learned indexes are competitive with traditional \btrees. Zhang et al.~\cite{DBLP:journals/pacmmod/ZhangSZ24} identify that directly extending learned indexes to disk environments often results in suboptimal performance compared to traditional \btrees. However, they show that employing optimizations such as leaf-page fetching strategies, prediction granularity, and adapting for disk characteristics can help bridge this gap.


Existing benchmarking studies involving learned indexes have not focused on external memory database joins, which are crucial for large-scale databases. 
External-memory joins offer vastly different trade-offs and challenges compared to in-memory joins. For relations stored in external memory, the dominating cost in performing the join is the I/O cost. Random I/Os are more expensive compared to sequential ones, resulting in different trade-off choices~\cite{YuanZJPACDKWBFJ17, JannenYZAEJMPRW15, Pandey0BBFJKP20}. Additionally, different storage devices such as hard disks and SSDs offer different I/O bandwidths and latencies, leading to a different choice of indexing parameters to achieve optimal performance. Therefore, the insights gained from employing learned models for accelerating in-memory joins are relevant but not directly applicable in the context of external memory.

\para{This paper} We investigate the applicability of learned indexes for external-memory joins, a common operation in large-scale databases. Through an extensive empirical evaluation, we assess whether the performance advantages of learned indexes in in-memory settings extend to external-memory environments. 

To utilize learned indexes for external-memory joins (\lj), we replace the traditional index in the indexed nested-loop join with a learned index and introduce optimizations tailored for external-memory settings to enhance performance. This approach aligns with the methodology used by Sabek and Kraska~\cite{sabek2023} in their study on in-memory learned joins. We conduct a comprehensive empirical evaluation, comparing \lj with various traditional external-memory join algorithms. Our study results in the following key contributions:

\begin{itemize}[leftmargin=*,noitemsep,nolistsep]
  \item We evaluate the performance of the indexed-nested loop join using learned indexes on real-world and synthetic sorted datasets.
  \item We evaluate learned indexes across a range of parameter settings, including table-size ratios (selectivity), concurrency, model accuracy, and types of storage devices (SSDs and HDDs).
  \item We evaluate the performance of learned indexes for joins on unsorted data to produce a sorted output by partitioning the data using the learned CDF model and evaluate its performance under memory pressure.
  \item We analyze experimental results and make several key observations that will benefit future work in incorporating learned indexes as part of the query execution engine in large-scale databases.
\end{itemize}

\para{Key Takeaways} Here we list the key takeaways from our experimental evaluation:
\begin{itemize}[leftmargin=*,noitemsep,nolistsep]
    \item For external-memory joins, learned indexes can benefit from using higher errors, and in turn achieve lower memory usage without loss in join performance. Using a larger error window size for the learned model helps increase disk-bandwidth utilization for learned indexes when the join no longer does sequential I/Os, which can happen when the sizes of two sorted tables being joined differ by a large factor. This enables smaller indexes and faster operations.        
    \item While learned indexes provide smaller index sizes and faster lookups, they do not reduce total I/O costs, leading to performance that is comparable to B-trees in external-memory joins.
    \item On SSDs, learned indexes improve the time to perform a join on sorted data compared to \btrees by $1.2-1.6\times$ when the input tables must be completely scanned. The improvement in performance on SSDs comes from faster CPU performance in querying for items. 
    \item Learned indexes scale similarly to \btrees with increasing threads, but they become I/O-bound at high levels of parallelism.
    \item Learned indexes have an order-of-magnitude higher build time than \btrees when built on the entire dataset. Sampling keys to build the learned index enables faster index construction without a noticeable impact on the join performance. However, even with sampling, the learned index build time is $10\times$ larger than \btrees.
\end{itemize}

\section{Background \& Related work}

In this section, we briefly describe the various learned index designs and their usage in database applications. 
We highlight the key insights and unanswered questions pertaining from these studies.

Kraska et al.~\cite{kraska_case_2018} showed that machine learning models can help reduce the memory footprint and increase the performance of traditional indexing data structures such as \btrees, hash tables, and filters.  More specific to databases, there is considerable interest in designing \emph{learned indexes} for sorted data that deliver better performance with lowered space usage than \btrees that are commonly used to implement database indexes. 

Various learned index designs have been proposed, including read-only indexes such as the Recursive model index (RMI)~\cite{kraska_case_2018}, RadixSpline~\cite{kipf_radixspline_2020} and Piecewise geometric model (PGM) index~\cite{ferragina_pgm-index_2020} and updatable learned indexes such as ALEX~\cite{ding_alex_2020}, FITing Tree~\cite{galakatos_fiting-tree_2019} and LIPP~\cite{wu_updatable_2021} as replacements for \btrees in static and dynamic workloads, respectively. For a more detailed history of the work in this area, please refer to~\cite{almamun2024survey} which extensively surveys learned indexes.  

There has also been considerable research showing the applicability of learned models in other applications. Recent works include improvements to fundamental algorithms such as sorting and joins ~\cite{DBLP:conf/sigmod/KristoVCMK20,sabek2023}, range index designs ~\cite{ding_alex_2020,wu_updatable_2021,ferragina_pgm-index_2020}, query performance in log-structured merge-trees~\cite{dai2020, abulibdeh2020learned}, multi-dimensional and spatial indexes~\cite{liLISALearnedIndex2020a, zhangSPRIGLearnedSpatial2021} and genome sequencing~\cite{DBLP:journals/bioinformatics/KirscheDS21, DBLP:journals/corr/abs-1910-04728}.

\subsection{Learned indexes} 

In this section, we briefly describe various types of learned indexes.

\para{Range indexes are CDF models} The key insight that ties machine learning techniques and deterministic problems such as indexing is that all range indexes model the cumulative distribution function (CDF) of the data they index ~\cite{kraska_case_2018}. The cumulative distribution function of a list is a function $F(x)$ that maps the probability that $x$ is greater than an item picked randomly from that list $L$. When the CDF value for an item $x$ is scaled by the number of items in the list, it returns the position where $x$ would fit in that list, i.e. $pos(x) = F(x).|L|$, where $|L|$ is the size of the list.

\begin{figure}
    \centering
    \includegraphics[page=1,scale=1.1]{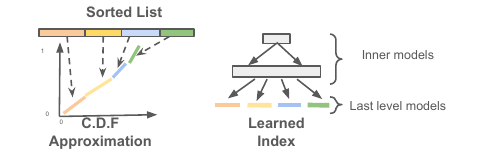}
    \caption{Modeling the CDF using piecewise linear approximation and indexing into the individual linear models.}
    \label{fig:learnedindex-datalayout}
    \vspace{-0.5cm}
\end{figure}


\para{CDF-based models}  Learned index implementations follow a common design trend in modelling the CDF. They often employ regression techniques (as opposed to deep neural networks) to model the CDF. 
As a single model is practically not enough to model the entire CDF, the CDF is modelled in a piecewise manner with a hierarchical structure indexing individual piecewise models. This is illustrated in \Cref{fig:learnedindex-datalayout}.

\begin{figure}
    \includegraphics[page=2,scale=1.1]{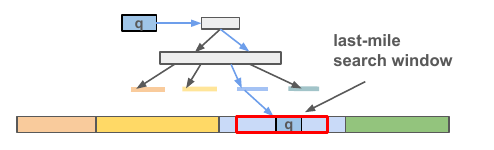}
    \caption{Query path using a learned index. The inner levels direct the query to the approximate location in the dataset. A last-mile search is performed to find the query key.}
    \label{fig:learnedindex-lookup}
    \vspace{-0.5cm}
\end{figure}

The CDF model only approximates the position of a key in the list, i.e., it cannot accurately determine the position of the key. However, this is not a significant issue in practice as the model error is usually bounded and a local \emph{last-mile search} can be performed to return the exact answer to a query (\Cref{fig:learnedindex-lookup}). The model error is a training parameter that trades accuracy for space usage, i.e., high-accuracy models require more space but a smaller last-mile search and vice versa. 


\subsection{Learned index designs}

Multiple learned index designs have been proposed, each using a unique modelling technique and data layout to implement the index. Below, we go over a few learned indexes that we use as candidates in our external-memory join implementation.

\para{Recursive model index (RMI)} RMI~\cite{kraska_case_2018} builds a hierarchy of models on the dataset given a specification by the user. The user specification provides various parameters, such as the type of models and learning techniques to use, the maximum branching factor of each level, and the maximum size of the final model. To help with tuning, the RMI implementation provides an optimizer that performs a grid search of RMI models against the space/performance curve, allowing users to choose the RMI configuration appropriate for their use case.
    
\para{RadixSpline} RadixSpline~\cite{kipf_radixspline_2020} is a single-pass learned index that uses a linear spline to model the CDF and employs a radix table to speed up the lookup of spline points. To perform a key search, the radix table is first consulted to return a range of linear spline segments. The returned linear spline segments are then used to find the error-bounded position of the lookup key. The RadixSpline is parametrizable in two parameters, the error-bound of the linear spline model and the number of the bits to use for the radix table. 
A larger error bound reduces the number of linear spline segments used to model the CDF but increases the last-mile search window. 
Similarly, using more bits for the radix table speeds up finding the correct linear spline model for a query at the cost of a larger radix table.
    
\para{Piecewise geometric model (PGM) index} The PGM index ~\cite{ferragina_pgm-index_2020} is a learned index that is similar to the RMI in that it builds a hierarchical structure of models. 
Each level is an error-bounded linear regression model built using piecewise linear approximation (PLA), i.e., the model is a piecewise function where each component is a linear function. 
The model can be visualized as a list of line segments that approximate a curve. 
The lowest level model approximates the CDF curve of the data, while higher levels are PLA models approximating the CDF curve of line-segment end points of the next level. Similar to the RadixSpline, the PGM is configurable by two parameters, the error bound and the maximum height of the index. 

\para{Adaptive Learned Index (ALEX)} 
The aforementioned learned indexes are read-only indexes. 
ALEX~\cite{ding_alex_2020} is an updatable learned index that augments the traditional \btree nodes with learned models. 
ALEX maintains the invariant that the data inside a node follows the distribution of the learned model associated with the node. 
ALEX uses a cost model to help with decisions related to rebalancing the tree when new items are inserted or deleted. 
While ALEX is implemented as an in-memory data structure, we use the implementation adapted for disks by Lan et al.~\cite{lan_updatable_2023}.

\subsection{Related work}
In this section, we describe related research work evaluating learned indexes.

\para{Benchmarks for learned indexes} The Search on Sorted Datasets (SOSD) benchmark~\cite{marcus2020a} evaluates the lookup performance of various learned indexes in main memory on sorted data. 
Lan et al. ~\cite{lan_updatable_2023} adapt multiple updatable learned indexes to external memory and evaluate performance on read, scan, and write workloads. 
They observe that in their current iterations, none of the updatable learned indexes are competitive with \btrees on all workloads.

\para{Learned indexes for in-memory joins and sorting}  Sabek and Kraska~\cite{sabek2023} show that learned indexes can improve the performance of in-memory joins.
They show that learned indexes must be carefully adapted for each of the three main types of joins --- hash, index-nested loop, and sort join --- and using them as black-box replacements for traditional indexes is not optimal. 
For instance, they observe that index-nested loop joins benefit from removing the last-mile search by redistributing elements according to the learned model predictions. 
As another example, hash and sort join benefit from using the sampled CDF to uniformly partition the workload across different cores. 
Similarly, Kristo et al.~\cite{DBLP:conf/sigmod/KristoVCMK20} show that utilizing CDF approximations to recursively partition unsorted data into buckets results in a sorting algorithm that is $1.49-5\times$ faster than various state-of-the-art sorting algorithms.

\section{Approach and Analysis}
\label{sec:approach}

In this section, we describe and analyze the usage of learned indexes for external-memory joins. We describe the overall approach of the learned index-based join and then describe specific optimizations. 
Finally, we analyze the performance in I/O cost using the affine memory model~\cite{dictAffine}. 

\subsection{Approach}\label{sec:approach}
\begin{figure*}
    \centering
    \includegraphics[page=7,scale=1.5]{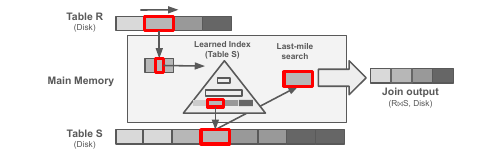}
    \caption{Illustration of the \lj: Tables $R$ and $S$ (with $S$ being the larger table) reside on disk, while the learned index for $S$ is kept in memory. Pages of table $R$ are sequentially loaded into memory, and for each key, the learned index predicts the corresponding page in $S$ to perform the final lookup.} 
    \label{fig:learnedjoin}
    \vspace{-0.5cm}
\end{figure*}

We use a learned index to support the indexed nested-loop join. During the join, keys from the smaller table are streamed, and each key is looked up in the learned index of the larger table to determine if it contributes to the join result (see~\Cref{fig:learnedjoin}). Since the join is performed using indexes, the resulting output is naturally sorted.

For sorted tables, the \lj uses two optimizations. The first optimization reduces the overall cost of the learned index look-up. The second optimization minimizes the size of the last mile search. We describe both optimizations below.

\para{Optimization 1}
The first optimization is to avoid traversing the entire height of a learned index for each query. 
Typically, a learned index query needs to start from the root and traverses down to the leaf that contains the model assigned to handle queries for the query key (\Cref{fig:learnedindex-lookup}). However, we can take advantage of the fact that queries increase monotonically during the join operation to reduce the query cost. Instead of traversing the entire height of the learned index for each query, we use a leaf node iterator that traverses the breadth of the last level. This iterator will advance to the next model when the query (which is the candidate join key) exceeds the range assigned to the model of the current leaf. 

\para{Optimization 2} The second optimization also takes advantage of the fact that queries during a join on sorted data occur in monotonically increasing order.
In other words, the search window only advances forward. For example, if the lower bound for key $K_i$ is at $L_i$, and the learned index returned a search window $[L_j, H_j]$ for key $K_j$ where $j > i$, then the last-mile search for key $(K_j)$ can be constrained to $[\text{MAX}(L_i, L_j), $ $H_j]$.

\subsection{Analysis}
We analyze the cost of using a learned index for indexed nested loop joins.

\para{Setup}  We perform the join on two input tables $R, S$ containing $|R|$ and $|S|$ number of keys respectively. We will assume that the tables are large enough such that their combined size exceeds main memory, and that a learned index for $S$ has already been built offline. 
We will also assume that the learned index fits in memory.
Without loss of generality, we will assume that $|R| < |S|$. 
For a query key $q$, the learned index of table $S$ will return a search window for where $q$ might exist. 
The size of the window will never exceed $\epsilon$, the maximum error bound of the learned index, and is a training parameter for the learned index.

\para{The affine model}  We analyze the performance of the \lj in the affine model~\cite{dictAffine}. Traditionally, the disk-access machine (DAM)~\cite{dam} model has been used to evaluate the performance of external-memory algorithms and data structures in terms of the number of I/O transfers in the memory hierarchy. However, the DAM model does not assign a cost to each I/O. On HDDs, it does not model the faster speeds of sequential I/O versus random I/O. On SSDs, it does not model internal device parallelism or the incremental cost of larger I/Os. 

The affine model ~\cite{ruemmlerIntroductionDiskDrive1994, andrewsNewAlgorithmsDisk1996, dictAffine} makes small refinements to the DAM model, but yields a surprisingly large improvement in predictability without sacrificing ease of use. The affine explicitly account for seeks (in spinning disks) by modelling the cost of an I/O of $k$ words as $1 + \alpha k$, where $\alpha \ll 1$ is a hardware parameter. 

\para{I/O cost analysis} We will divide the I/O cost of the learned index into two components, one for reading $R$, and the other for reading $S$. 
Assuming $R$ is read in blocks of size $B$, the block transfer size - the cost of reading $R$ is $\frac{|R|}{B}$. 
For $S$, we assume that the learned index-based join will perform a single I/O of $\epsilon$ words for each query. The cost of I/O is then $O(|R|(1 + \alpha \frac{\epsilon}{B}))$, where $\alpha$ is the hardware parameter according to the affine model. This cost is also bounded by $O(\frac{|S|}{\epsilon}(1 + \alpha \frac{\epsilon}{B}))$ as we do not read any item in $S$ more than once. 
Hence, the I/O cost for reading $S$ is $O(\min(|R|,\frac{|S|}{\epsilon})(1 + \alpha \frac{\epsilon}{B}))$. 
The overall I/O cost can now be summarized as 
\begin{equation}
    O \left( \frac{|R|}{B} +  \min(|R|,\frac{|S|}{\epsilon})\cdot{}(1 + \alpha \frac{\epsilon}{B}) \right)
\end{equation}

In the case of $|R| \le \frac{|S|}{\epsilon}$, this analysis shows that increasing the size of the last-mile search window by building less accurate learned indexes does not significantly increase the overall I/O cost as $\alpha << 1$. 
When the table sizes are very similar $(|R| \ge \frac{|S|}{\epsilon})$, the I/O cost of the learned index is essentially the same as linearly scanning both tables. 
In this case, the affine model predicts that increasing $\epsilon$ actually decreases the I/O cost - we perform fewer but larger I/Os to completely scan $S$. Note that this analysis also holds for unsorted tables, the only difference being that the indexes are unclustered.

\para{Index size analysis} While there are no tight bounds proven for the size of learned indexes, empirically they occupy less space than \btrees. Analyzing the size of learned indexes is not straightforward as it depends on the distribution of the data, the specific design of the learned index, and the training parameters used to build the learned index. 
The most critical training parameter common to all learned index designs is the size of the last-mile search window. 
Ferragina et al.~\cite{ferragina_performance_2021} find that when the gaps between elements follow a distribution with finite mean and variance, the size of the learned index is $O(n/\epsilon^2)$, where $\epsilon$ is the size of the search window and $n$ is the number of keys. 

In general, we can assume that learned indexes occupy less space than \btrees, and that the space decreases by more than a linear factor with the accuracy of the learned index. 
This is in contrast to \btrees whose size decreases linearly with the node size. 
We will assume that the size of the learned index is $O(n/f(\epsilon))$, where $f$ is a function that is polynomially greater than a linear function.

\subsection{Takeaways}

We now compare the cost of using the learned index compared to ithe \btree in the indexed nested-loop join (INLJ). We then compare the \lj proposed above with the sort join. We compare these in terms of I/O cost, CPU cost, and index size.

\para{Learned index vs \btrees in INLJ} Both indexes have the same I/O cost for the same search window size ($\epsilon$ for learned indexes, node size for \btrees). Increasing $\epsilon$ or the node size increases the I/O cost marginally ($\alpha << 1$) for both the indexes. However, learned indexes occupy less memory and are faster to query compared to \btrees. The learned index also has a better memory performance trade-off curve. Although the size of both the learned index and \btree can be reduced by increasing the size of the search window, the size of the learned index decreases by more than a linear factor of $\epsilon$ compared to the \btree size that only decreases linearly with the node size.

\para{Learned index-based INLJ vs sort join} The comparison between the sort join and \lj is similar to comparing the sort join with INLJ. 
The I/O cost for both methods depends on the size of the input tables. 
When the tables are of similar size, the I/O cost of both the join methods are equal. 
In this case, the differentiating factor of both the join methods is the CPU cost. 
The cost of the sort join is one key comparison per element from the larger table, resulting in a total CPU cost of $O(|S|)$. 
For the index-based approaches, the CPU cost is one index lookup per element from the smaller table, leading to a total CPU cost of  $O(|R|\cdot\log{|S|})$. When table sizes vary by a factor of more than $\epsilon$ (that is, $|R| \le \frac{|S|}{\epsilon}$), the \lj, similar to the index nested-loop join, costs less I/O and CPU compared to the sort join.

\section{Implementation}
We now describe our implementation of join methods. We first describe how our tables are stored on disk, followed by the join and merge implementations. We then describe our join implementation for unsorted data. In both cases, the tables are stored on disk.

\subsection{Tables}

While fully featured database systems will use more complex data storage layouts and robust page management strategies, we intentionally choose a very simplistic layout to focus on the effects of learned indexes for joins. 

\para{Storage Layout} A table is a list of key-value tuples. The keys and values are of a fixed size of 8 byte keys and 8 byte values. The keys are distinct in a table. Tables are stored as a dense sorted array on disk in a single file. The first 16 bytes of the file are reserved for a header to store the number and size of the key-value tuples, followed by the key-value tuples themselves. We do not perform any compression of the actual key-value tuples.

\para{Reading and writing tables}\label{sec:read} Tables are logically divided into blocks, each 4096 bytes in size. To read a key-value tuple from the table, the block corresponding to that key must be loaded into memory. The blocks are stored in an internal buffer that holds a contiguous set of blocks in main memory. 
We configure the internal buffer size to hold enough blocks so that the search window of the learned index can be fully held in memory. 
If a key that we are searching for is already in the internal buffer, we immediately return the key. 
Otherwise, the set of contiguous blocks starting with the block containing the key is loaded into memory. Similarly, when writing the output join table, we store keys in an internal buffer and flush them to disk once the buffer is full.

\begin{figure*}[h]
    \centering
    \includegraphics[page=8, scale=1.4]{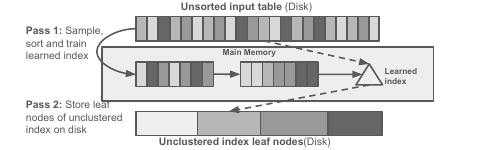}
    \caption{An unclustered index using a learned index is built in two passes: the first pass samples the data to train the learned index, and the second pass uses the index to assign each data item to its predicted position.}
    \label{fig:unclustered}
    \vspace{-0.5cm}
\end{figure*}

\subsection{Indexes} 

Indexes are built offline and serialized to disk. When used for a join, the indexes are loaded from disk and stored fully in main memory. 

\para{Index API} Given a query for a key, all indexes (learned or \btrees) return a search window whose size is bounded, representing the range in which the table contains the lower bound of the query key, which is the largest key in the table that is greater than or equal to the query key.
More formally, a table $T$ is an array of $[K_1, K_2, ... K_N]$, such that $K_i > K_{i-1}$. The index is a function $I_T(x)$ that returns the pair $(lo, hi)$ such that $K_{lo} \ge x \le K_{hi} $ and $(hi-lo) \le \epsilon$, where $\epsilon$ is the error bound of the learned index.


\para{\btree index} We use the interpretation defined in~\cite{kraska_case_2018}, where the inner nodes of the \btree are interpreted as a learned model and the leaf nodes are the search window ranges that these learned models ultimately return. When interpreted this way, the \btree can be decomposed into two distinct entities - a learned index (inner nodes) and the data (leaf nodes). For the sorted tables, we only create the learned index part of the \btree by evenly sampling keys (these are keys that would have been the first key of their leaf nodes) and insert them into a \btree. 

\para{Learned indexes} We consider various learned index implementations (as described in~\Cref{sec:exp_setup}) in our evaluation. We ensure that all learned indexes use the same search window size (except RMI, where this is not configurable). We discuss the exact configuration used for each learned index in the experimental setup (\Cref{sec:exp_setup}).

\para{Constructing learned indexes using sampling}
Constructing learned indexes on sorted arrays can be expensive due to the computational cost of modeling the data distribution. In contrast, building a \btree is more efficient, as pivots can be directly selected from the sorted array.

To reduce the cost of learned index construction, we sample every $k^{\text{th}}$ element from the array and learn the distribution over this subset. We then build a learned index on the sampled array with error $\epsilon'$. A query returning the interval $[lo, hi]$ in the sampled array maps to the interval $[k \cdot lo,\ k \cdot hi]$ in the full array. This effectively gives a learned index on the original array with error $\epsilon = k \cdot \epsilon'$, constructed using only $n/k$ elements.

\para{Learned Index Lookup} Using learned indexes to answer \emph{lower\_bound} queries is a two-step process. First, a lookup is performed on the index for the query key $Q$ in table $T$. The index will then return a range $[lo, hi]$, representing the range containing the actual lower bound. The second step is to perform a last mile search in the table to find the exact lower bound in the table. 


\para{Last-mile search} 
To perform the actual last-mile search, we use the branchless implementation of the binary search, resulting in a fast and efficient last-mile search. 
The branchless search algorithm takes advantage of the conditional move (CMOV) instruction to generate assembly code with no branches~\cite{cite-branchless-search}.
We first check if the search range returned by the index is partially loaded in our internal buffer.  
If the lower bound is not found or the internal buffer contains blocks that do not overlap with the search range, we load the required contiguous set of blocks from the disk into memory.

\subsection{Join on sorted tables}

We now describe implementation details of the various join implementations. The join operation will output the common keys from the input tables ($R, S$) as its own table to disk. We compare the indexed nested loop join with \btree and \lj, against each other and also with other join methods such as hash join and sort join. 

\para{Indexed nested loop join} Similar to the hash join, we start first by streaming all keys of $R$. For each key of $R$, we query the index of $S$ which will return a search window. We then do a last-mile search inside $S[lo, hi]$, outputting the key to the final table if it is present. 

\para{Sort join} If $R, S$ are already sorted, we skip the sort phase and proceed directly to the merge phase. We use a standard two-pointer approach to find the common keys. We initialize two iterators, comparing the iterator heads and advancing the iterator with the smaller key. If the keys are equal, the key is added to the output join table. The join ends if any of the iterators reaches the end of its table and cannot advance.

\para{Hash join} We first iterate over all keys of $R$ and insert them into an in-memory hash table $H_R$. 
We use \texttt{std::unordered\_map} as our in-memory hash table. We then iterate over all the elements from $S$, adding the key to the output if it exists in $H_R$.

\subsection{Join on unsorted tables}

We now describe the implementation details of the join methods for unsorted tables. These methods will output the result of the join in sorted order.
Here we only implement two variants of the indexed nested loop join, \btree based and learned index based. We build unclustered indexes for both input tables as part of the join.

\para{\btree index} We first build a complete \btree for each input table by sequentially inserting keys along with a pointer to its location in the table. We then perform a range scan on the \btree of the smaller table to stream the keys in sorted order, checking if each key exists in the large table by querying the \btree of the larger table. The keys which are common to both input tables are added to join table.

\para{Learned index} To approximate the data distribution, we build a learned index on a set of sampled keys from the unsorted table. 
The learned index on the sampled subset acts as a model for the distribution of data in the entire dataset, as uniform sampling captures the overall data distribution and also avoids the need to fully sort the data.

We use the approximate rank of a key in the sampled learned index to partition all keys into the desired number of partitions. Each key $k$ is assigned to the partition $\frac{\hat{R_k}}{N}P$, where $\hat{R_k}$ is the approximate rank of $k$, $N$ is the number of keys in the table, and $P$ is the desired number of partitions. We choose $P$ so that the average partition fits into a small constant number of pages. In practice, if the keys were randomly sampled uniformly and the learned index is reasonably accurate, the partition sizes have low variance. 
As long as the $\hat{R_k}$ returned by the learned index is monotonic, the partitions will be disjoint. The partitions are essentially leaf nodes of an unclustered index, and the partitioning process is illustrated in \Cref{fig:unclustered}.

Partitioning the data according to the position returned by the CDF model is an idea that has been explored by Kristo et al.~\cite{DBLP:conf/sigmod/KristoVCMK20} in the context of in-memory sorting, where the data is sorted by recursively partitioning the data using the learned index. 
Similarly, Abu-Libdeh et al.~\cite{abulibdeh2020learned} build tables that are sorted into blocks according to the position predicted by the learned index. 
We adapt this idea for joins on external memory by only partitioning the data once and not sorting the data inside the partitions. Our partitioning method uses only two passes on the data, one to sample the keys and build the CDF and the other to assign the key to a bucket according to the built model. 

After partitioning keys into buckets, we perform the indexed nested-loop join. We store the learned index and the partition map of both tables in memory. We sequentially load each bucket of $R$ into memory and process the keys in sorted order. For each key, we query the index of $S$ and load the corresponding bucket into memory. The key is added to the output if it is part of both tables.

\section{Evaluation}
\begin{table}[h]
\resizebox{\columnwidth}{!}{ 
  \begin{tabular}{|l|l|l|l|}
    \toprule
    Dataset &  Size  & Key count & Description    \\
    \midrule
    FB      & 3.2 GB & 200000000  & Facebook user ids \\
    Wiki    & 1.44 GB & 90437011 & Wikipedia edit timestamps \\
    OSM     & 12.8 GB & 800000000 & OpenStreetMap locations\\
    Books   & 12.8 GB & 800000000 & Amazon book popularity data.\\    
    \midrule
    udense      & 3.2 GB & 200000000 &  sequential integers \\
    usparse      & 3.2 GB & 200000000 & Uniform sparse distribution \\
    normal      & 3.2 GB & 200000000  & Normal distribution \\
    lognormal      & 3.2 GB & 200000000 & Lognormal distribution \\
    \bottomrule
  \end{tabular}
  }
  \caption{Summary of datasets}
  \label{tab:datasets}

\end{table}
\begin{figure*}
  \begin{subfigure}{\linewidth}
    \centering
    \includegraphics[page=36]{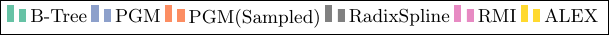}
  \end{subfigure}
  
  \begin{subfigure}[T]{0.35\linewidth}
    \centering
    \includegraphics[page=35]{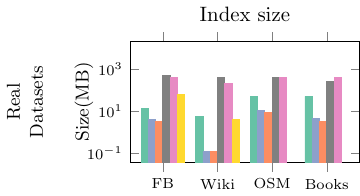}
  \end{subfigure}
  \begin{subfigure}[T]{0.30\linewidth}
    \centering
    \includegraphics[page=34]{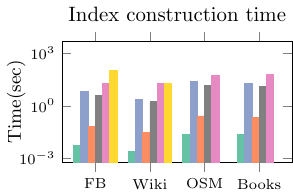}
  \end{subfigure}
  \begin{subfigure}[T]{0.30\linewidth}
    \centering
    \includegraphics[page=37]{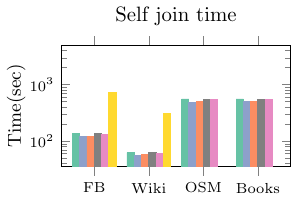}
  \end{subfigure}
  
  \begin{subfigure}[T]{0.35\linewidth}
    \centering
    \includegraphics[page=39]{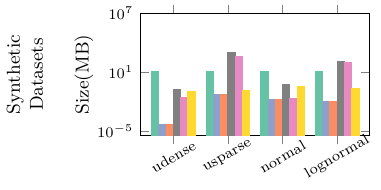}
  \end{subfigure}
  \begin{subfigure}[T]{0.30\linewidth}
    \centering
    \includegraphics[page=38]{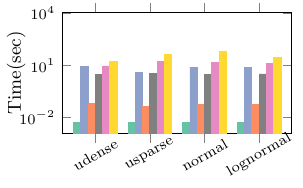}
  \end{subfigure}
  \begin{subfigure}[T]{0.30\linewidth}
    \centering
    \includegraphics[page=58]
    {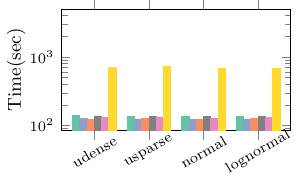}
  \end{subfigure}
  \caption{Index size, build time and performance of learned indexes and \btree. Index Size is the space used by the index in main memory. ALEX did not finish building the index using 32GB of RAM for OSM/Books datasets.}
  \label{fig:index}
  \vspace{-0.5cm}
\end{figure*}


In this section, we evaluate the usage of learned indexes for external-memory joins against traditional join algorithms. For \lj, we employ the learned model to replace the index in indexed nested loop joins. For all indexes, we keep the leaf nodes on disk and the non-leaf nodes in main memory. First, we evaluate various learned indexes on construction time, in-memory space requirements and join performance when used as part of \lj. 
We then pick the most appropriate learned index and compare the \lj against traditional indexed nested loop joins, sort join (SJ) and hash join (HJ). 

We evaluate the performance of the \lj against traditional join algorithms across multiple dimensions: (1) storage device types (HDD/SSD), (2) data ordering (sorted/unsorted), (3) concurrency, (4) constrained memory settings and (5) trade off between error window size and number of threads.  

All benchmarking source code and datasets used in our evaluation are available at \url{https://github.com/saltsystemslab/learnedjoindiskexp}.


\subsection{Experimental Setup}\label{sec:exp_setup}
In this section, we describe the experimental setup we use for our evaluation.


\para{Environment} We run our experiments on an Intel(R) Core(TM) i7-8700 CPU @ 3.20GHz with 1 NUMA node with 12 cores with a single 12MB L3 cache with 32GB of RAM. We run our experiments on both SSD and HDD. The SSD model used in our experiments is a 512GB SanDisk SD9SB8W5, while the HDD is 2TB TOSHIBA DT01ACA200.

\para{Datasets} We use real-world and synthetic datasets used in the unified benchmarking paper by Marcus et al.~\cite{marcus2020a}.\Cref{tab:datasets} summarizes various datasets.

\begin{itemize} [leftmargin=*,noitemsep,nolistsep]
    \item \textbf{Real-world datasets:}  \textbf{Books} is a collection of 800 million keys representing book popularity data from amazon. \textbf{Wiki} is a collection of 90 million wikipedia edit timestamps. \textbf{OSM} is a collection of 800 million OpenStreetMap locations. \textbf{FB} is a collection of 200 million Facebook user ids. 
    
    \item \textbf{Synthetic datasets:} All synthetic datasets contain 200M items generated from a universe of 64-bit unsigned integers. \textbf{usparse} represents a dataset of integers picked uniform randomly, while \textbf{udense} represents a dense distribution of sequential keys. \textbf{normal} and \textbf{lognormal} represent data distributions that follow normal and lognormal distributions, respectively.
\end{itemize}

\para{Join processing} Each join operation is invoked as a new process. Our join evaluation setup for \textbf{static tables} is similar to previous join studies~\cite{albutiu2012a, balkesen2013, bandle2021, blanas_design_2011, thostrup, sabek2023, krastnikov_efficient_2020}. The indexes are constructed offline, stored on disk, and loaded into memory before starting the join operation. For multithreaded experiments, the input tables are partitioned evenly across threads and work is distributed evenly. All operations are performed with \textbf{cold cache} by dropping the operating system page caches, and the input files are read using \texttt{O\_DIRECT} to ensure that all data is always retrieved from disk. We do not use \texttt{O\_DIRECT} for writes as the output cost for all join methods is the same and increases the duration of the experiment. To constrain the memory during join process, we employ \texttt{CGroupsV2} utility.

\subsection{Evaluating learned indexes for joins on disk} \label{sec:micro}


In this section, we evaluate the performance of PGM, RadixSpline, RMI, and ALEX against \btrees for index construction, query latency, and in-memory index space. We use the insights from this evaluation to identify the appropriate learned index to employ for external-memory joins on sorted and unsorted data.

Our evaluation of learned indexes extends the SOSD benchmark~\cite{marcus2020a} for disk-based evaluation and performance on the join operation. Furthermore, our evaluation extends the disk-based benchmarking study~\cite{lan_updatable_2023} by evaluating the learned indexes on external memory joins.

\para{Indexes}  We use the \btree index for implementing the index nested-loop join. We use the STX-BTree v0.9~\cite{bingmann_stx_2023} library as our \btree implementation. For each of the learned indexes (PGM~\cite{vinciguerra_gvinciguerrapgm-index_2023}, RadixSpline~\cite{kipf_radixspline_2020} and the recursive model index (RMI)~\cite{kraska_case_2018}), we use the reference implementations provided by the authors. For ALEX~\cite{ding_alex_2020}, we use the disk-based implementation used in the study by Lan et al.~\cite{lan_updatable_2023}, which evaluates updatable learned indexes in external memory. The specific configuration used for building each index is described below:

\begin{itemize}[leftmargin=*,noitemsep,nolistsep]
  \item \textbf{\btree}: \btree node size is fixed to 4K bytes. To make the comparison fair with static learned indexes, the \btree is built by bulk loading the keys. Furthermore, we build the \btree on every $256^{\text{th}}$ key in the dataset. 
  The leaf nodes map their keys to the block they come from in the dataset, where a block is a contiguous set of 256 keys. 
  Thus, the leaf nodes of a \btree return a search window of size 256, similar to how a learned index returns a search window for the last-mile search.
  \item \textbf{Piecewise geometric model (PGM}): The PGM index is built with an error window of 256 with a single level. 
  \item \textbf{Piecewise geometric model (PGM) sampled}:  A version of the PGM index that is built on every $128^{\text{th}}$ key with a maximum error of 2. The search window returned by this index is then scaled up to get the search window in the dataset.
  \item \textbf{RadixSpline (RS)}: 
  For each dataset, we choose the \emph{Pareto-optimal} index configuration of the RadixSpline as evaluated by the SOSD benchmark~\cite{marcus2020a}. The \emph{Pareto-optimal} index is an index configuration for which no other index with lower memory usage has better performance. Radix bits range from 16 to 28 across the various datasets. We also set the RadixSpline maximum error to 256. 
  \item  \textbf{Recursive model index}: 
  For each dataset, we use the RMI model configuration from the SOSD benchmark ~\cite{marcus2020a} with hyperparameters tuned using CDFShop ~\cite{marcusCDFShopExploringOptimizing2020}. Similar to the RadixSpline, we choose the \emph{Pareto-optimal} index for each dataset. 
  \item  \textbf{ALEX}: We use the disk-based implementation provided by Lan et al. ~\cite{lan_updatable_2023} to evaluate the performance of updatable learned indexes on disk. The data nodes are stored as a contiguous file on disk, while the inner nodes are stored in main memory.
\end{itemize}

\para{Setup} The construction time is measured as the time to build the index over the set of keys. We load all the keys into memory before building the index to avoid counting the disk I/O cost during index construction. All indexes, except RMI, can be built in a single pass by streaming keys from the disk. 
The query performance is evaluated by measuring the time taken to perform the index nested-loop self-join using the index. All indexes are loaded into main memory while the data is kept on disk and loaded one page at a time. 
The index evaluation experiments are constructed on flash storage with the operating system page caches flushed before the start of each experiment. 
\Cref{fig:index} plots the construction time, index size, and the time to complete the self join on all datasets using each index.

\para{Index size} 
%
For real-world datasets, PGM indexes (both sampled and full) have the lowest memory footprint among all learned indexes, being $4\times$ smaller than the \btree. The RadixSpline and RMI are an order of magnitude ($30-80\times$) larger than \btrees. ALEX was unable to construct the \emph{OSM} and \emph{Books} datasets as it ran out of memory. For synthetic datasets, PGM indexes were an order of magnitude smaller in size compared to \btrees. The RMI and RadixSpline are not able to model the \emph{usparse} and \emph{lognormal} distributions well and needed several orders of magnitude of space to do so.
%
%

\para{Index construction time} All learned indexes take at least 4 orders of magnitude longer to build than \btrees. Constructing the PGM index with sampling reduces the duration by roughly two orders of magnitude. The higher construction times of learned indexes can be reduced by sampling without sacrificing query performance. 
As the time to construct the PGM and RadixSpline indexes grows linearly with the size of the dataset, evenly sampling keys to construct the index reduces the number of keys used to train the index while still effectively capturing the distribution of the data. 
The PGM index built on the sampled data has performance identical to that of the index built on the entire dataset. 
Additionally, the PGM index is also simpler to construct, requiring only a single parameter (the maximum error), compared to the RMI and RadixSpline, which require tuning multiple parameters to find the pareto-optimal configuration for space and performance. 

Despite reducing the construction time using sampling, learned indexes are slower to construct than \btrees. 
The \btree construction is extremely fast when built using bulk loading because the \btree only needs to perform fixed memory allocations and data copying. 
On the other hand, learned indexes need to perform more complex processing to model the distribution leading to increased construction time.

\para{Join performance}
%
Across all datasets, we find that the PGM index performs the most consistently, being $1.1-1.7\times$ faster than the \btree. 
As disk I/O is the bottleneck, all learned indexes (except ALEX) performed very similarly.
For ALEX, our results are consistent with the disk-resident learned index study by Lan et al. ~\cite{lan_updatable_2023}, which showed that the read performance of ALEX on disk is not competitive in read-heavy workloads. 

\para{Takeaways}  Overall, the PGM index with sampling offers the best tradeoff in terms of query latency, construction time, and space usage. Using sampling, the join can be sped up by $1.1-1.7\times$ compared to \btrees and also uses $4\times$ less space. Although the PGM index takes $10\times$ longer to build compared to the \btree, this is often an acceptable tradeoff in large-scale analytics systems where the indexes are built offline and are used several times to perform fast joins. \textbf{Therefore, we employ sampled PGM index in our implementation of the \lj.}

\begin{figure*}[h]
  \centering
  \begin{subfigure}{\linewidth}
    \centering
    \includegraphics[page=3]{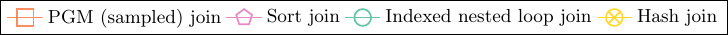}
  \end{subfigure}

  \centering
    \centering
    \begin{subfigure}{0.24\linewidth}
      \centering
        \includegraphics[page=4]{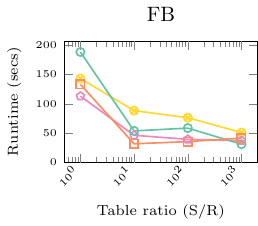}
    \end{subfigure}
    \centering
    \begin{subfigure}{0.24\linewidth}
      \centering
        \includegraphics[page=41]{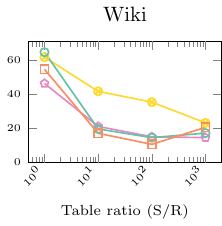}
    \end{subfigure}
    \centering
    \begin{subfigure}{0.24\linewidth}
      \centering
        \includegraphics[page=54]{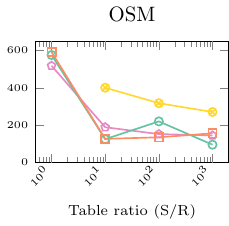}
    \end{subfigure}
    \centering
    \begin{subfigure}{0.24\linewidth}
      \centering
        \includegraphics[page=55]{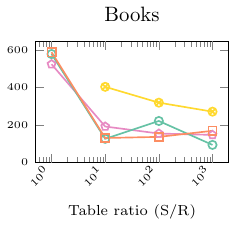}
    \end{subfigure}

    \centering
    \begin{subfigure}{0.24\linewidth}
      \centering
        \includegraphics[page=42]{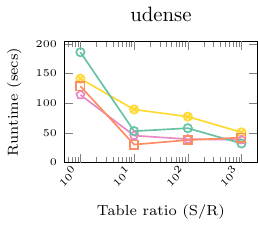}
    \end{subfigure}
    \centering
    \begin{subfigure}{0.24\linewidth}
      \centering
        \includegraphics[page=43]{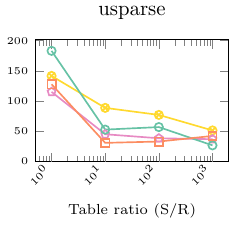}
    \end{subfigure}
    \centering
    \begin{subfigure}{0.24\linewidth}
      \centering
        \includegraphics[page=44]{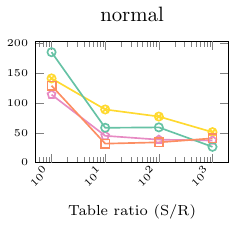}
    \end{subfigure}
    \centering
    \begin{subfigure}{0.24\linewidth}
      \centering
        \includegraphics[page=45]{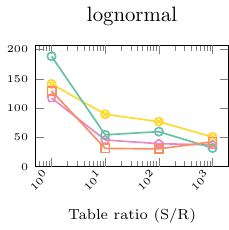}
    \end{subfigure}
    \caption{Performance of various join methods for sorted tables on flash-based storage devices (SSD). }
    \label{fig:ssd}
\end{figure*}

\begin{figure*}
  \centering
  \begin{subfigure}{\linewidth}
    \centering
    \begin{subfigure}{0.24\linewidth}
      \centering
        \includegraphics[page=5]{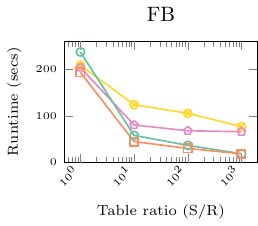}
    \end{subfigure}
    \centering
    \begin{subfigure}{0.24\linewidth}
      \centering
        \includegraphics[page=46]{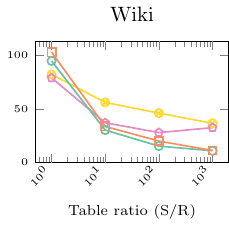}
    \end{subfigure}
    \centering
    \begin{subfigure}{0.24\linewidth}
      \centering
        \includegraphics[page=56]{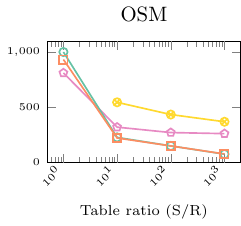}
    \end{subfigure}
    \centering
    \begin{subfigure}{0.24\linewidth}
      \centering
        \includegraphics[page=57]{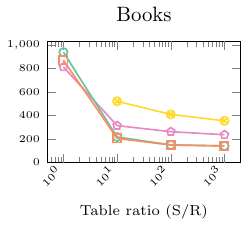}
    \end{subfigure}
    
    \begin{subfigure}{0.24\linewidth}
      \centering
        \includegraphics[page=47]{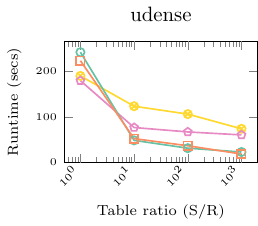}
    \end{subfigure}
    \centering
    \begin{subfigure}{0.24\linewidth}
      \centering
        \includegraphics[page=48]{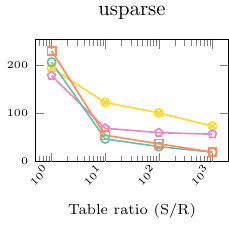}
    \end{subfigure}
    \centering
    \begin{subfigure}{0.24\linewidth}
      \centering
        \includegraphics[page=49]{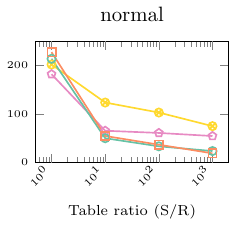}
    \end{subfigure}
    \centering
    \begin{subfigure}{0.24\linewidth}
      \centering
        \includegraphics[page=50]{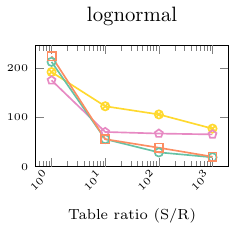}
    \end{subfigure}
  \end{subfigure}
    \caption{Performance of various join methods for sorted tables on hard disks (HDD). }
    \label{fig:hdd}
    \vspace{-0.5cm}
\end{figure*}

\subsection{Join methods on sorted data}
In this section, we compare the single-threaded performance of the hash join (HJ), indexed nested-loop join (INLJ), sort join (SJ), and \lj on sorted tables stored on disk (both HDD and SSD) across varying table size ratios.

\para{Setup} We use the sampled PGM index as the index for the \lj based on the analysis presented in~\Cref{sec:micro}. The indexed nested-loop join uses a \btree as the index. The sorting phase of sort join is skipped as the data is already sorted on disk. The hash join uses STL \texttt{std::unordered\_map} as the hash table of the smaller table. All indexes and hash tables are loaded in memory before starting the join, while the data is streamed from disk in pages using file I/O. The join time experiments on SSD and HDD is plotted in~\Cref{fig:ssd} and~\Cref{fig:hdd}, respectively. 

\para{Join selectivity} We employ different table ratios to evaluate join algorithms for different selectivity values. A table size ratio of 1 is a self-join. For other table size ratios (10, 100, and 1000), we sample a fraction of the keys uniformly randomly from the table to create the smaller table for the join. 

\para{Index for indexed-nested loop join}  
We employ \defn{sampled PGM index} in our implementation of the \lj. This is based on the conclusions drawn from an extensive study of learned indexes for construction time, size, and query time detailed in~\Cref{sec:micro}.
A sampled version of the PGM index is built on every $128^{\text{th}}$ key with a max error of 2. The search window returned by this index is scaled up to get the actual search window in the dataset.
We use the \btree index to implement the index-nested loop join. We use the STX-BTree v0.9~\cite{bingmann_stx_2023} library as our \btree implementation.

\para{Join method evaluation on SSDs} The \lj is faster by $1.2-1.6\times$ compared to the indexed nested-loop join with \btree 
when the table size ratio lies between 1 and 100. 
The \lj is also faster compared to the sort-join($1.1-1.4\times$) when the table size ratio is between 10 and 100. 
When the table size ratio is between 1 and 100, the I/O cost is identical for all methods as items from the table are fetched in blocks of size 4KB that contain 256 items. 
Thus, every block is expected to contain a join key. 
At higher table size ratios (such as 1000), both indexed-based joins, learned and \btree based, perform random I/Os on the inner table. In our tests, the \lj performed slightly worse than the \btree based index nested-loop join by about ($1.1-1.6\times$). 
However, the performance is still very similar in absolute terms due to the small output size. 
\textbf{In cases where both tables have to be scanned completely, the \lj offers better performance compared to the \btree by $1.2-1.6\times$, and the sort join by $1.1-1.4\times$ (except when both input tables are of similar size)}.


\para{Join method evaluation with HDDs} On hard disks, the performance of the \btree and \lj was similar across all table-size ratios and datasets. Both the index-based methods were also faster than the sort-join except for when the table sizes were equal. \textbf{Indexed-based methods have similar performance on HDDs and are faster than sort-join except for when table sizes are similar.}

\para{Takeaways} Learned indexes offer performance mostly similar to the traditional \btree-based index nested-loop join in external memory settings. 
The PGM index is much smaller in size compared to the \btree. 
However, that does not result in improved performance as the join operation is bounded by the disk I/O. 
Using learned indexes for join does not help in reducing the total I/O. 
This is especially true when there is enough working memory to store the \btree. 
This is unlike in main-memory joins where learned indexes can help speed up join performance~\cite{sabek2023}.


\begin{figure}
  \centering
  \begin{subfigure}{\linewidth}
    \centering
    \includegraphics[page=51]{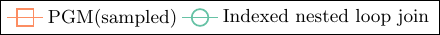}
  \end{subfigure}

  \centering
  \begin{subfigure}{\linewidth}
    \centering
    \begin{subfigure}{0.49\linewidth}
      \centering
        \includegraphics[page=52]{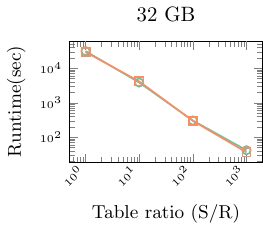}
    \end{subfigure}
    \begin{subfigure}{0.49\linewidth}
      \centering
        \includegraphics[page=59]{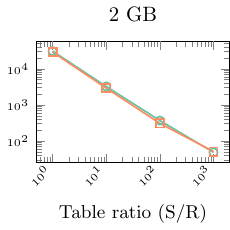}
    \end{subfigure}
  \caption{Join duration on unsorted data after index creation under different memory constaints using the FB dataset.}
  \end{subfigure}
  \label{fig:unsorted}
  
  \begin{subfigure}{\linewidth}
  \centering
    \begin{tabular}{|l|c|c|}
        \toprule
            & \multicolumn{2}{c|}{Index creation (sec)}\\
            \midrule
            Memory Limit & \btree & PGM \\
            \midrule 
            2GB   &  10777.06 & 2993.49 \\
            32GB  &  1406.33 &  411.28 \\
            \bottomrule
    \end{tabular}
  \caption{Time to create the index on unsorted data under memory constraints using the FB dataset.} 
  \label{tab:unsorted-partition}
  \end{subfigure}
  \vspace{-0.5cm}
  \caption{Join performance on unsorted data}
  \vspace{-0.5cm}
\end{figure}

\subsection{Join methods on unsorted data}
In this section, we compare the single-threaded performance of various joins on unsorted tables to produce a sorted join output in external memory.

\para{Setup} We generate input tables for a dataset by shuffling keys from the FB dataset and storing them on disk. 
We compute the join using an indexed nested-loop join using unclustered indexes (\btree, PGM) on the keys. 
The index stores keys and a pointer to its table entry. 
We use the pointer to fetch the associated value of a join key from the table on disk. 
We run the test under different memory constraints using \texttt{CGroupsV2} to limit the amount of memory that a process is allowed to use and plot the time to complete the join for the \btree and PGM index in~\Cref{fig:unsorted}. Similar to joins on sorted data, we test for different table size ratios of the input tables.
We summarize the index construction and implementation of the join for each index.

\begin{itemize} [leftmargin=*,noitemsep,nolistsep]
    \item \textbf{\btree}: We build the \btree by streaming keys from disk. The leaf nodes of the \btree are stored on disk, while the intermediate nodes are held in memory. Nodes are 4KB in size. To compute the join, we scan the keys from the smaller table and for each key perform a lookup in the \btree of the larger table. 
    
    \item \textbf{PGM Index}:  As the tables are not sorted, we partition the data into disjoint ranges with the help of a PGM index built on a sampled subset (1\%) of the dataset. 
    The rank of a key according to the sampled PGM index is used to determine its partition. 
    As the rank returned is only an approximation, it is not necessary that partitions are equally sized.
    The more accurate the learned index, the less the variance in the size of the partitions. We set the expected partition size to be that of a single page (4KB).
    Partitions are flushed to disk 8 keys at a time to improve write efficiency. 
    The index and partition map for each table are stored in main memory, and during the join only a single partition per table is kept in memory. 
    To compute the join, we sequentially process the partitions of the smaller table.
    For each key in the partition, we use the PGM index of the larger table to determine the corresponding partition in the larger table. Our approach for learned joins on unsorted data is based on learned sorting~\cite{DBLP:conf/sigmod/KristoVCMK20}.
\end{itemize}

\para{Results summary}
The performance of different join approaches is largely similar. Joins on unsorted data incur more I/O operations than those on sorted tables. Since I/O cost dominates the join performance, faster queries of the \lj do not yield significant improvements. 
The PGM index also performs similarly to the \btree when memory is constrained using \texttt{CGroupsV2}. 
Although the PGM index is significantly smaller at 3MB compared to 12MB for \btree, this 75\% reduction in index size is insufficient to reduce the I/O performed to swap pages to disk due to constrained memory.
However, partitioning the data to partially sort the data using the PGM index is faster ($3-3.5\times$) compared to constructing a \btree with random inserts. 
This is due to the higher write amplification of \btree to keep items sorted under random inserts.  

\para{Takeaways} Although partitioning and constructing an unclustered index for a unsorted table using the CDF model is upto $3.5\times$ faster compared to building the \btree index, the join itself performs similar. The benefits of smaller indexes and faster queries are not apparent, as the memory savings of the \lj are only over the inner nodes of \btree, which is tiny compared to the space required to store the dataset. \textbf{The benefits of smaller indexes and faster queries on the join itself are not apparent even when operating under constrained memory settings}.

\begin{figure*}
      \centering
      \begin{subfigure}{\linewidth}
        \centering
        \includegraphics[page=3]{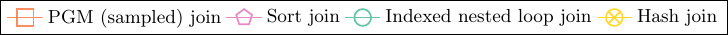}
    \end{subfigure}

      \centering
      \begin{subfigure}{0.24\linewidth}
        \centering
        \includegraphics[page=6]{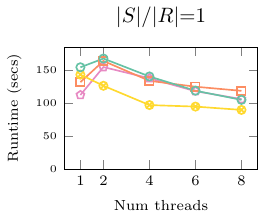}
    \end{subfigure}
    \centering
    \begin{subfigure}{0.24\linewidth}
        \centering
        \includegraphics[page=7]{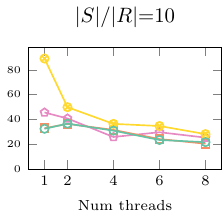}
    \end{subfigure}
    \centering
    \begin{subfigure}{0.24\linewidth}
        \centering
        \includegraphics[page=8]{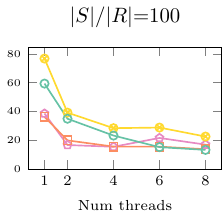}
    \end{subfigure}
    \centering
    \begin{subfigure}{0.24\linewidth}
        \centering
        \includegraphics[page=9]{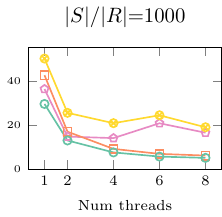}
    \end{subfigure}
    \caption{Performance of join methods as the number of threads increase (FB dataset)}
    \label{fig:joinThreadStudy}
  \end{figure*}
\begin{table}
\centering
  \begin{tabular}{|l|r|r|r|}
    \toprule
& \multicolumn{3}{c|}{Index size (MB)}\\
\midrule 
    Dataset & $\epsilon=256$ & $\epsilon=2048$ & $\epsilon=4096$ \\
    \midrule
    FB & 3.1623 & 0.7314 & 0.0910 \\
    Wiki & 0.1121 & 0.0507 & 0.0064 \\
    OSM & 8.6735 & 2.4492 & 0.6454 \\
    Books & 3.140 & 0.092 & 0.024 \\
    \midrule
    udense & 48 KB & 48 KB & 48 KB\\
    usparse & 0.0505 & 0.0034 & 0.0002 \\
    normal & 0.0109 & 0.0054 & 0.0027 \\
    lognormal & 0.0152 & 0.0076 & 0.0038 \\
    \bottomrule
  \end{tabular}
  \caption{Index size of PGM index as search window size varies.}
  \label{tab:indexsize}
  \vspace{-0.5cm}
\end{table}

\subsection{Multithreading} In this section, we evaluate the effect of scaling up external-memory join methods using multiple threads. We will further study the performance tradeoff between search window size and number of threads.

\para{Setup} We partition the smaller table into equal size partitions based on the number of threads and assign a single partition to each thread. 
We build an index on the larger table and query it for keys from the smaller table to find a match for the join. 
Each thread writes its output to a separate file on disk. 
The threads are synchronized to block until all threads finish writing their join output. Once all threads are finished, we merge the output file for the final join output. 
To do this, each thread computes the offset of where its output lies in the merged output and writes it to the final join output. 
We test for thread sizes of 1, 2, 4, 6, and 8. 
We compare the \lj with the indexed nested-loop join with \btree, hash join, and sort join and plot the results in ~\Cref{fig:joinThreadStudy}. 



\para{Results summary} Adding more threads makes the join operation more I/O bound.  Both the \lj and \btree indexed joins scale linearly with increasing threads before becoming I/O bound at some point. For example, when the table size ratio is 100, the \lj becomes completely I/O bound with 4 threads. At this point, adding more threads does not improve the overall process as the join is I/O bound. The performance of sort join does not scale with more threads as it is already I/O bound. The hash join is mostly CPU bound and almost linearly scales with increasing threads all the way up to 8 threads. The runtime does not include the time to build the hash table. Thus, the time for hash join measurement avoids the time to perform I/O on the smaller table. This makes the hash join less I/O bound compared to the other joins.  Note that the hash join uses much more memory compared to indexed and sort joins as it stores a hash table of size $O(|R|)$ in memory. It is only included in the evaluation only as a baseline for comparison.

\para{Takeaway} 
The conclusions made in the previous section regarding which join method to use at different table ratios hold true even for multithreaded join processing.
\textbf{Learned indexes scale up with more threads similarly compared to other join methods.} 

\begin{figure*}
    \begin{subfigure}{\linewidth}
        \centering
        \includegraphics[page=11]{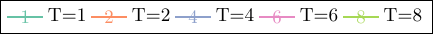}
    \end{subfigure}
    
    \centering
    \begin{subfigure}{0.24\linewidth}
        \centering
        \includegraphics[page=12]{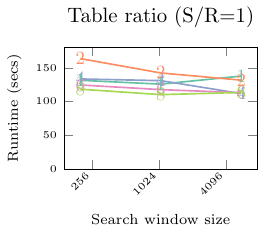}
    \end{subfigure}
    \centering
    \begin{subfigure}{0.24\linewidth}
        \centering
        \includegraphics[page=13]{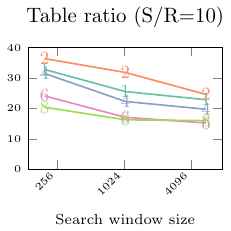}
    \end{subfigure}
    \centering
    \begin{subfigure}{0.24\linewidth}
        \centering
        \includegraphics[page=14]{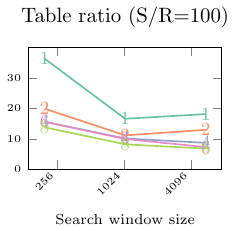}
    \end{subfigure}
    \centering
    \begin{subfigure}{0.24\linewidth}
        \centering
        \includegraphics[page=15]{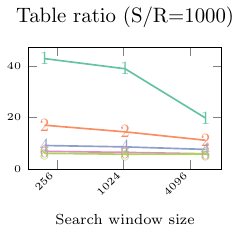}
     \end{subfigure}

    \caption{Performance of \lj with various search window sizes (FB dataset)}
    \label{fig:joinEpsilonThread}
    \vspace{-0.5cm}
\end{figure*}

\subsection{Error window size analysis} In this section, we study the effect of the size of the error window on the join performance with an increasing number of threads. We also study how the build time and index size varies as the error window changes.

\para{Setup}  We use the sampled PGM index with error window sizes of 256, 1024 and 4096. The results of this experiment are plotted in~\Cref{fig:joinEpsilonThread}, while \Cref{tab:indexsize} shows the index size of the sampled PGM. We increase the page fetch size of a single I/O call to match the size of the error window. Note that this does not change the total number of bytes fetched from disk, only the number of I/O calls performed to fetch those bytes.

\para{Result Summary} For the sampled PGM index, the index size is reduced by a factor of $30-130\times$ when the error window size is increased from 256 to 2048 and 4096 respectively. 
The index build time is independent of the index error window size and depends only on the sampling rate (which is fixed at 128 in this case). 


\Cref{fig:joinEpsilonThread} shows that the performance of the PGM index remains consistent with increasing the size of the search window. We perform I/Os of larger block sizes to ensure that we perform no more than a single I/O call per query. For lower table size ratios, performance remains consistent with increasing the search window size from 256 to 4096. At these table size ratios, the join disk access pattern is sequential. Thus, requesting larger I/O block sizes across a varying number of threads does not have a significant effect on overall performance. As the table size ratios increase, the join access pattern is no longer sequential. When run with a low number of threads, a larger search window and I/O block fetch sizes lead to higher disk bandwidth utilization, resulting in disk saturation and consequently better performance. With a high number of threads, the disk utilization is already high and performing larger I/O block fetches has no effect on performance.

\section{Conclusion}
\vspace{-0.1cm}

This study presents an extensive evaluation of learned indexes for external-memory joins, analyzing their impact across varoius database configurations. 
Unlike the main-memory setting, where learned indexes provide clear advantages, our findings indicate that their benefits in external-memory joins are less pronounced due to I/O dominance.

While learned indexes offer smaller index sizes and faster lookups, they do not reduce the total I/O costs, resulting in similar performance to B-trees-based joins in most cases. However, by tuning parameters such as search window size and error bounds, learned indexes can achieve comparable or slightly better performance in specific workloads, particularly on SSDs. The significant index construction overhead (up to $1000\times$ slower than B-trees) further limits their practicality for dynamic workloads, but remains acceptable in offline analytics scenarios.

Our results suggest that practitioners must carefully assess I/O constraints when integrating learned indexes into database engines. 

\section*{Acknowledgments}
This research is funded in part by
NSF grant OAC 2339521 and 2517201.

\balance
\bibliographystyle{plain}
\bibliography{bibliography}
\end{document}